\documentclass[11pt]{article}

\usepackage{amssymb,latexsym, amsmath}
\usepackage{graphicx}

\makeatletter
\@addtoreset{figure}{section}
\def\thefigure{\thesection.\@arabic\c@figure} \def\fps@figure{h, t}
\@addtoreset{table}{bsection}
\def\thetable{\thesection.\@arabic\c@table} \def\fps@table{h, t}
\@addtoreset{equation}{section}

\makeatother

\advance \topmargin by -\headheight
\advance \topmargin by -\headsep
\evensidemargin \oddsidemargin
\newtheorem{thm}{Theorem}[section]
\newtheorem{prop}[thm]{Proposition}

\newfont{\tenbi}{cmbxti10}

\def\one{\mathbf 1}

\def\D{\mathcal D}

\def\V{\mathcal V}

\def\H{\mathcal H}

\def\L{\mathfrak L}

\def\n{\mathfrak n}
\def\b{\mathfrak b}
\def\O{\mathfrak O}

\def\B{\mathbf B}

\def\R{\mathbb R}

\begin{document}

\title{Poisson brackets on rational functions and multi-Hamiltonian structure
for integrable lattices}

\author {Leonid Faybusovich 
\thanks{Research partially supported by the NSF grant  DMS98-03191}
\\
Department of Mathematics\\
University of Notre Dame\\
Notre Dame, IN 46556 \\
{\footnotesize Leonid.Faybusovich.1@nd.edu}\\
\and
Michael Gekhtman
\thanks{Corresponding author}
\\
Department of Mathematics\\
University of Notre Dame\\
Notre Dame, IN 46556 \\
{\footnotesize Michael.Gekhtman.1@nd.edu}
}

\maketitle

\begin{abstract}
We introduce a family of compatible Poisson brackets
on the space of rational functions with denominator of a fixed
degree and use it to derive a multi-Hamiltonian structure for
a family of integrable lattice equations that includes
both the standard and the relativistic Toda lattices.
\end{abstract}
\medskip

\medskip

\medskip

\section{} 
It has been known since Moser's work on finite non-periodic 
Toda lattice (\cite{moser}) that the space of rational functions
plays an important role in solving integrable lattice and constructing
action-angle variables. The Moser map that sends the space of $n\times n$ tri-diagonal
matrices (the phase space for the Toda lattice) into a space of proper
rational functions with a denominator of degree $n$ was later utilized in a more general
context (see, e.g. \cite{dlnt}, \cite{b}, \cite{f}) and generalized in \cite{k}
for an arbitrary semisimple Lie algebra to be used, mainly, as a tool for linearization
for a class of finite systems of differential-difference equations.

The goal of this paper is to show that  the Moser map also serves as an
effective tool in establishing a multi-Hamiltonian nature of a class
of integrable lattices that includes, in particular, the standard and the
relativistic Toda lattices. This class was studied in our recent paper
\cite{fg}. Our approach is in contrast to those of \cite{dam1}, \cite{dam2},
\cite{sur2}, where the search for compatible Poisson structures was conducted
in terms of the matrix entries of the Lax operator. Instead, we introduce
explicitly the family of compatible Poisson brackets on the space of rational
functions and then pull them back to a phase space of a lattice in question
via the Moser map. Note that the linear Poisson structure on the phase space
then corresponds to the Atiyah-Hitchin Poisson structure on rational functions
\cite{at}. (In the case of the finite non-periodic  Toda lattice, this
correspondence was first explicitly pointed out in \cite{f}.)

In sect. 2 we review the construction of the family of integrable
systems introduced in \cite{fg} as restrictions of the full Kostant-Toda
flows in $sl(n)$ to elements of a certain family of symplectic leaves.
A construction of compatible Poisson structures on rational functions and a 
consequent description of multi-Hamiltonian structure for lattices of sect. 2
is given in sect. 3. In the last section we give explicit formulae in 
terms of the corresponding Lax operators
for master symmetries that generate these multi-Hamiltonian structure. In 
the case of the symmetric Toda lattice such master symmetries were implicitly
defined via recursive relations in \cite{dam1}. 

We conclude this introduction with two natural questions that
we would like to address in the future. First, it would be interesting to extend
our approach to more general (e.g. generic)
symplectic leaves of the full Kostant-Toda flows. Second, we would like to understand 
how results of this paper can be translated into a geometric language for bi-Hamiltonian
systems advocated in \cite{gz}.

We would like to thank Yu. Suris and referees for valuable remarks.

\medskip
\section{}

Let us first recall the definition of the Kostant-Toda flows.
Denote by $e_j\ (j=0,\ldots, n)$ vectors of the standard basis in $\R^{n+1}$, by
$E_{ij}$ an elementary matrix $(\delta_i^\alpha \delta_j^\beta)_{\alpha, \beta=0}^n$ and by 
$J$ an
$(n+1)\times (n+1)$ matrix with
$1$s on the first sub-diagonal and
$0$s everywhere else. Let $\b_+, \n_+, \b_-, \n_-$ be, resp., 
algebras of upper triangular, strictly upper triangular, lower triangular and strictly lower
triangular $(n+1)\times (n+1)$ matrices. Denote by $\H$ the set $J + \b_+$ of upper Hessenberg
matrices.

For any matrix $A$ we write its decomposition into a sum of lower triangular and strictly upper triangular matrices as
$$
A=A_- + A_0 + A_+ 
$$
and define $A_{\ge 0}= A_0 + A_+,A_{\le 0}= A_0 + A_-$.

The hierarchy
of commuting Kostant-Toda flows on $\H$ is given by the family of  Lax equations
\begin{equation}
\dot X = [ X, (X^m)_{\le 0}]\  (m= 1,\ldots, n).
\label{lax}
\end{equation}
Each of the flows defined by (\ref{lax}) is Hamiltonian with respect to 
a linear Poisson structure on $\H$ is 
obtained as a pull-back of the Kirillov-Kostant
structure on $\b_-^*$, the dual of $\b_-$, if one identifies $\b_-^*$
and $\H$ via the trace form. 
Then a Poisson bracket of two functions $f_1, f_2$ on
$\H$ is
\begin{equation}
\{f_1, f_2\} (X)=
\langle X, [(\nabla f_1 (X))_{\le 0}, (\nabla f_2 (X))_{\le 0}] \rangle\ ,
\label{brack1}
\end{equation}
where we denote by $\langle X, Y\rangle$ the trace form $\mbox{Trace}(XY)$
and gradients are computed w.r.t. this form.
The $m$-th flow of the hierarchy (\ref{lax}) is generated by the Hamiltonian
\begin{equation}
H_{m+1}(X)=\frac{1}{m+1} Tr(X^{m+1})= \frac{1}{m+1} \sum_{i=0}^{n} \lambda_i^{m+1}\ ,
\label{hamk}
\end{equation}
where $\lambda_i (i=0,\ldots, n) $ are the eigenvalues of $X$.

{\it The Weyl function}
\begin{equation}
M(\lambda)=M(\lambda,X)=((\lambda\one-X)^{-1} e_0,e_0)=
\sum_{j=0}^\infty\frac{s_j(X)}{\lambda^{j+1}}
\label{weyl}
\end{equation}
is an important tool in the study of the Toda flows (\ref{lax})
(see, e.g., \cite{b}, \cite{bf}, \cite{dlnt},\cite{f}, \cite{moser}).
Here $s_j(X)= (X^{j} e_0,e_0)$. 

If
$X\in\H_0$, where $\H_0\subset\H$ consists of elements with simple real
spectrum $\lambda_1<\ldots< \lambda_n$,
one can write (\ref{weyl}) as
\begin{equation}
{\displaystyle
M(\lambda)=\sum_{i=0}^n \frac{\rho_i(X)}{\lambda -\lambda_i(X)}\ 
\qquad ( \sum_{i=0}^n \rho_i(X) = 1 )\ .
}
\label{weylreg}
\end{equation}

The Lax equation (\ref{lax}) implies the following evolution for
$ \rho_i(X), \lambda_i(X)$
\begin{equation}
\dot\rho_i(X)=(\lambda_i(X)^m-s_m) \rho_i(X), \ \dot\lambda_i(X)=0\ .
\label{evolrho}
\end{equation}
An identity
\begin{equation}
s_m= \sum_{j=0}^n \lambda_j^m  \rho_j(X) \
\label{moment}
\end{equation}
allows one to write the system (\ref{evolrho}) in a closed form.
The solution of (\ref{evolrho}) is given by
\begin{equation}
{\displaystyle
\rho_i(X(t))=\frac{e^{\lambda_i^m t} \rho_i(X(0))}
{\sum_{j=0}^n e^{\lambda_j^m t} \rho_j(X(0))}\ .
}
\label{rhot}
\end{equation}

As is well-known, symplectic leaves of the bracket (\ref{brack1}) are
orbits of the coadjoint action of the group $\B_-$ of lower triangular invertible matrices:
\begin{equation}
\O_{X}=\left \{ J + (\mbox{Ad}_n X)_{\ge 0}\ : n\in \B_-\right \}\ .
\label{coad}
\end{equation}

In \cite{fg} we described a family of integrable lattices associated with
orbits $\O_{X}$ of a special kind. This family contains both the standard
and relativistic Toda lattices. Its members are parameterized by increasing sequences of natural
numbers
$I=\{i_1,\ldots,i_k\ : 0< i_1< \ldots< i_k=n\}$. To each sequence $I$ there
corresponds a $1$-parameter family of $2 n$-dimensional coadjoint
orbits $M_I=\O_{X_I+ \nu \one}\in \H$, where 
\begin{equation}
{\displaystyle
X_I=e_{0 i_1} + \sum_{j=1}^{k-1}e_{i_j i_{j+1}}\ .
}
\label{X_I}
\end{equation}

Two Darboux parametrizations for  $M_I$ were found in \cite{fg}. Each of them
allows us to lift the first ($m=1$) of the Kostant-Toda flows (\ref{lax}) on
$M_I$ to an integrable flow on $\mathbb R^{2n+2}$ equipped with the standard
symplectic structure. In one of these parametrizations the Toda Hamiltonian
$H_1(X)$ has a form
\begin{equation}
{\displaystyle
\tilde H_I(Q,P) =\frac{1}{2} \sum_{i=0}^n P^2_i\quad + \sum_{0\le i < n; i\ne
i_1,\ldots,i_{k-1}}^n P_i e^{Q_{i+1} - Q_i}\ +\
\sum_{j=1}^{k-1}e^{Q_{i_j+1} - Q_{i_j}}\ .
}
\label{hamIexp}
\end{equation}

The set $M'_I$ of elements of the form
\begin{equation}
{\displaystyle
X=(J+D) (\one - C_k)^{-1} (\one - C_{k-1})^{-1}\cdots (\one -
C_1)^{-1},
}
\label{factorI}
\end{equation}
where $D=\mbox{diag} (d_0,\ldots, d_n)$
\begin{equation}
{\displaystyle
C_j= \sum_{\alpha=i_{j-1}}^{i_j-1} c_\alpha e_{\alpha,\alpha+1}\ ,
}
\label{Cj}
\end{equation}
is dense in $M_I$. 

Then the first ($m=1$) of the Kostant-Toda flows (\ref{lax})
on $M'_I$ is equivalent to the following system
\begin{eqnarray}
\nonumber
&\dot d_i=d_i(c_i -c_{i-1}),\\
\nonumber
&\dot c_i=c_i(d_{i+1}-d_i + c_{i+1}-c_{i-1})\qquad  (i_j < i<i_{j+1}, j=0,\ldots, k-1) \\
&\dot c_{i_j}=c_{i_j} \left ( d_{i_j+1}-d_{i_j} + (1-\delta_{i_j+1,i_{j+1}}) c_{i_j+1})\qquad
(j=0,\ldots, k-1)\ .
\right )\ 
\label{rtlI}
\end{eqnarray}
Equations (\ref{rtlI}) can be viewed as particular cases of the 
{\it constrained KP lattice} introduced and discretized in \cite{sur3}.
In fact, all the minimal indecomposable invariant submanifolds for the latter
system can be obtained this way. To emphasize a connection to Toda flows and
the dependence on $I$ we will denote the lattice (\ref{rtlI}) by $TL_I$.
If $I=\{ n\}$, we recover the relativistic Toda lattice 
\begin{equation}
\dot c_i=c_i(d_{i+1}-d_i + c_{i+1}-c_{i-1}),\ \dot d_i=d_i(c_i - c_{i-1})\ .
\label{rtl}
\end{equation}
If, on the other hand, $I=\{1,2,\ldots, n\}$, then $M_I$ is the set  $Jac$ of tri-diagonal
matrices in $\H$ with non-zero entries on the super-diagonal and thus, one obtains the standard
Toda lattice. In coordinates $c_i, d_i$ equations of motion form a system
\begin{equation}
\dot d_i=d_i(c_i -c_{i-1}),\ \dot c_i=c_i(d_{i+1}-d_i), 
\label{volt}
\end{equation}
which, after relabeling $d_i=u_{2i-1}, c_i=u_{2i}$, becomes the Volterra lattice
$$
\dot u_i=u_i(u_{i+1}-u_{i-1}).
$$

In \cite{fg} we proved the following

\begin{prop}
For any $I$, there exists a unique birational transformation of the from
$X \to Ad_{n(X)} X$
from $M_I$ to $Jac$, that preserves the Weyl function  $M(\lambda, X)$
and, for $k=1,\ldots, n$, sends the $k$-th Toda flow (\ref{lax})
on  $M_I$ into the $k$-th Toda flow on $Jac$. Here $n(X)$ is a unipotent
upper triangular matrix, whose off-diagonal elements in the first row are 
all equal to zero.
\label{propfg}
\end{prop}

One of the consequences of Proposition \ref{propfg} is that, for any $I$,
on the open dense set in $M_I$, an element $X\in M_I$ can be uniquely
determined by its  Weyl function  $M(\lambda, X)$. (This, of course, is well-known
in the case of the tri-diagonal and relativistic Toda lattices, see, e.g.,
\cite{moser}, \cite{b}, \cite{bf}, \cite{kmz}.) In the next section, we use this
fact to derive the multi-Hamiltonian structure for systems $TL_I$.

\medskip
\section{}

Let $Rat_{n+1}$ denote a space of rational functions of the form
\begin{equation}
{\displaystyle
m(\lambda)=\frac{q(\lambda)}{p(\lambda)}=\sum_{i=0}^\infty \frac{h_i}{\lambda^{i+1}}\ ,
\label{rat}
}
\end{equation}
where $p(\lambda)$ is a monic polynomial of degree
$n+1$ and $q(\lambda)$ is a polynomial of degree less than $n+1$.
To define a Poisson bracket on $Rat_{n+1}$, it is sufficient to specify pairwise brackets
for $p(\lambda), q(\lambda), p(\mu), q(\mu)$, where $\lambda$ and $\mu$ are arbitrary.

For fixed $p(\lambda), q(\lambda)$ and $k=0,\ldots, n$, let us denote
\begin{equation}
q^{[k]}(\lambda)=\lambda^k q(\lambda) \quad (\mbox{mod}\  p(\lambda))\ 
\label{q^{[k]}}
\end{equation}
and define a skew-symmetric bracket $\{\ ,\ \}_k$ on coefficients of $p(\lambda), q(\lambda)$
by setting
\begin{equation}
\{p(\lambda), p(\mu)\}_k = \{q(\lambda), q(\mu)\}_k =0
\label{brackpq1}
\end{equation}
and
\begin{equation}
{\displaystyle
\{p(\lambda), q(\mu)\}_k = \frac{p(\lambda) q^{[k]}(\mu) - p(\mu) q^{[k]}(\lambda)}{\lambda-\mu}\ .
}
\label{brackpq2}
\end{equation}

\begin{prop}
$\{\ ,\ \}_k$ ($k=0,\ldots, n$) are compatible Poisson structures on
$Rat_{n+1}$.
\label{prop1}
\end{prop}

{\bf Proof.} 
It is sufficient to check the statement on an open
dense subset of $Rat_{n+1}$ defined by the assumption that $p(\lambda)$ and $q(\lambda)$
are co-prime and all roots $\lambda_0,\ldots \lambda_n$ of $p(\lambda)$ are distinct.
On this subset 
\begin{equation}
\nonumber
{\displaystyle
m(\lambda)=\frac{q(\lambda)}{p(\lambda)}= \sum_{i=0}^n \frac{r_i}{\lambda -\lambda_i}
}
\end{equation}
and the data $\{ \lambda_i, q(\lambda_i), i=0,\ldots, n\}$ determines 
$p(\lambda)$ and $q(\lambda)$ completely. 
In particular,
\begin{equation}
{\displaystyle
r_k= q(\lambda_i) \prod_{j\ne i} \frac{1}{\lambda_i-\lambda_j}= \frac{
q(\lambda_i)}{p'(\lambda_i)}\  }
\label{rqlambda}
\end{equation}

By (\ref{brackpq1}), 
\begin{equation}
\{\lambda_i, \lambda_j\}_k  =0
\label{bracklambda}
\end{equation}

Since
$$ 
0=\{ p(\lambda_i), q(\mu)\}_k= p'(\lambda_i) \{\lambda_i, q(\mu)\}_k + 
\{p(\lambda), q(\mu)\}_k\upharpoonright_{\lambda=\lambda_i}, 
$$
one obtains from (\ref{brackpq2}) 
\begin{equation}
{\displaystyle
\{\lambda_i, q(\mu)\}_k =\frac{p(\mu) q^{[k]}(\lambda_i)}{p'(\lambda_i) (\lambda_i-\mu)}
= \frac{p(\mu)\lambda_i^k q(\lambda_i)}{p'(\lambda_i) (\lambda_i-\mu)}
=-\lambda_i^k q(\lambda_i) \prod_{j\ne i} \frac{\mu -\lambda_j}{\lambda_i-\lambda_j}\ ,
}
\label{bracklambdaqmu}
\end{equation}
which,  together with (\ref{bracklambda}), implies
\begin{equation}
{\displaystyle
\{\lambda_i, q(\lambda_j)\}_k = -\lambda_i^k q(\lambda_i) \delta_i^j,
}
\label{bracklambdaq}
\end{equation}
and, consequently,
\begin{equation}
\{q(\lambda_i), q(\lambda_j)\}_k  = 0\ .
\label{brackq}
\end{equation}

It follows from (\ref{bracklambda}), (\ref{bracklambdaq}), (\ref{brackq}), that
in coordinates $\lambda_i, q_i=q(\lambda_i)$ , 
any linear combination $\{ \ , \ \}_c = \sum_{k=0}^n c_k \{\ , \ \}_k$ has a form
\begin{eqnarray}
\nonumber
&\{\lambda_i, \lambda_j\}_c=\{q_i, q_j\}_c = 0\ ,  \\
&\{\lambda_i, q_j\}_c = -c(\lambda_i) q_i \delta_i^j\ ,
\label{brackc}
\end{eqnarray}
where $c(\lambda)= \sum_{k=0}^n c_k \lambda^k$.
It is easy to see that the bracket defined by (\ref{brackc}) satisfies
the Jacobi identity, with canonical coordinates given by
\begin{equation}
{\displaystyle
x_i = \int \frac{\mbox{d} \lambda_i}{c(\lambda_i)}\ , \ y_i=\ln q_i \ (i=0,\ldots, n).
}
\label{darboux1}
\end{equation}
Thus, any linear combination of $\{\ ,\ \}_k$ is a Poisson bracket, which
finishes the proof.
\hfill$\Box$

\medskip
\noindent{\bf Remarks. 1.} The expression in the right hand side of
(\ref{brackpq2}) is called {\it a Bezoutian} of polynomials $p(\lambda)$
and $q^{[k]}(\lambda)$. For more information on bezoutians and the role they in 
the control theory we refer the reader to a survey \cite{fh}

{\bf 2.} When $k=0$, brackets (\ref{brackpq1}), (\ref{brackpq2}) give Atiyah-Hitchin Poisson
structure on $Rat_{n+1}$ \cite{at}.
\medskip

Poisson structure (\ref{brackpq1}), (\ref{brackpq2}) can be re-written
directly in terms of the elements $m(\lambda) \in Rat_{n+1}$ as follows
\begin{equation}
{\displaystyle
\{m(\lambda), m(\mu)\}_k = \left ( (\lambda^k m(\lambda))_-
- (\mu^k m(\mu))_-\right )
\frac{m(\lambda)  - m(\mu)
}{\lambda-\mu}\ , }
\label{brackm}
\end{equation}
where, for a rational function $r(\lambda)$, $(r(\lambda))_+$ denotes
the polynomial part of its Laurent decomposition and 
$r(\lambda)_- =r(\lambda) - (r(\lambda))_+$. It follows from (\ref{brackm}) that
, in terms of coefficients $h_i$ of the Laurent expansion (\ref{rat}) of $m(\lambda)$,
$\{\ ,\ \}_k$ has a form
\begin{equation}
{\displaystyle
\{ h_i, h_j\}_k =\sum_{\alpha=i}^j h_{k+\alpha}h_{i+j-1-\alpha} \ \ (i<j)
}
\label{brackmoment}
\end{equation}

Now we can restrict brackets $\{\ ,\ \}_k$ to a subset of $Rat_{n+1}$
that contains the image of the map $M(\lambda, \cdot): \H \to Rat_{n+1}$
described by (\ref{weyl}). This subset, denoted by $Rat_{n+1}'$ is the set of
all $M(\lambda)=\frac{q(\lambda)}{p(\lambda)}\in Rat_{n+1} $ with both polynomials
$q(\lambda)$ and  $p(\lambda)$ monic.
To compute  a Poisson bracket induced by $\{\ ,\ \}_k$ on $Rat_{n+1}'$, we first
note that, by (\ref{brackpq2}), the bracket between $h_0$, the leading coefficient
of $q(\lambda)$,  and $p(\lambda)$  is $\{p(\lambda),\ h_0\}_k=q^{[k]}(\lambda)$.
Then, since $M(\lambda)=\frac{1}{h_0} m(\lambda)$ defines a surjective map from 
$Rat_{n+1}$ to $Rat_{n+1}'$, a straightforward computation leads to the following

\begin{prop}
The family of  compatible Poisson structures on
$Rat_{n+1}'$ is given by
\begin{equation}
{\displaystyle
\{M(\lambda), M(\mu)\}_k = \left ( (\lambda^k M(\lambda))_-
- (\mu^k M(\mu))_-\right )
\left( \frac{M(\lambda)  - M(\mu)
}{\lambda-\mu}\   + M(\lambda) M(\mu)\right ) \ .
}
\label{brackM}
\end{equation}
\label{prop2}
\end{prop}

Any linear combination $\{ \ , \ \}_c = \sum_{k=0}^n c_k \{ \ , \ \}_k$ of
brackets (\ref{brackM}) is degenerate. Indeed, it follows from 
(\ref{bracklambdaqmu}), that 
\begin{equation}
\{\lambda_i,\ h_0\}_k=-\frac{\lambda_i^k q_i}{p'(\lambda_i)}\ .
\label{qlead}
\end{equation}
Let $F=\sum_{i=0}^n x_i$, where $x_i$ are defined in (\ref{darboux1}).
Then, clearly,  $\{p(\lambda), F\}_c= 0$ and
$${\displaystyle
\{ \frac{q(\lambda)}{h_0}, F\}_c = \sum_{i=0}^n q(\lambda_i) \prod_{j\ne i}
\frac{\lambda -\lambda_j}{\lambda_i-\lambda_j} - \frac{q(\lambda)}{h_0}
\sum_{i=0}^n \frac{q(\lambda_i)}{h_0 p'(\lambda_i)}
}
$$
But, by (\ref{rqlambda}), $\sum_{i=0}^n \frac{q(\lambda_i)}{h_0 p'(\lambda_i)}=1$ and so
$\{ \frac{q(\lambda)}{h_0}, F\}_c =0$ by the Lagrange interpolation formula.
Thus, $F=\sum_{i=0}^n x_i$ is a Casimir for $\{ \ , \ \}_c$ on $Rat'_{n+1}$, while
canonical coordinates coordinates for this bracket can be easily derived from  (\ref{darboux1}):
\begin{equation}
{\displaystyle
x_i = \int \frac{\mbox{d} \lambda_i}{c(\lambda_i)}\ , \ y_i=\ln \frac{q_i}{q_n} \ (i=0,\ldots, n-1)
. }
\label{darboux2}
\end{equation}

Denote $\rho_i=\frac{r_i}{h_0}=\frac{q_i}{h_0 p'(\lambda_i)}$ 
and define $H_j= \frac{1}{j}\sum_{l=0}^n \lambda_l^j (j=\pm 1, \pm 2,\ldots )$ and 
$H_0= \sum_{l=0}^n \ln \lambda_l$.  Then (\ref{bracklambdaq}), (\ref{qlead})
imply
\begin{equation}
\{\rho_i, H_j\}_k= (\lambda_i^{k+j-1} -\sum_{l=0}^n \lambda_l^{k+j-1} \rho_l)\rho_i\ .
\nonumber
\end{equation}
Comparing the last equation with (\ref{evolrho}), (\ref{moment}), one sees that
equations of motion induced on $Rat'_{n+1}$ via the map (\ref{weyl}) by the $m$-th
Toda flow (\ref{lax}) coincide with Hamilton equations generated in the Poisson
structure $\{\ , \ \}_k$ by the Hamiltonian $H_{m+1-k}$. Now, by Proposition \ref{propfg},
for any index set $I$, (\ref{weyl}) defines an almost everywhere invertible map
from $M_I$ to $Rat'_{n+1}$. Moreover, the bracket (\ref{brackM}) is polynomial in
terms of the coefficients of the Laurent expansion $M(\lambda, X)= \sum_0^\infty h_j
\lambda^{-j-1}= \sum_0^\infty (L^j e_0, e_0)
\lambda^{-j-1}$, the well-known determinantal formulae, expressing the entries
of the element of $Jac$  via $h_j$ are rational in  $h_j$ (\cite{akh}), and,
by Proposition \ref{propfg}, so are formulae expressing matrix entries of elements
of $M_I$ in terms of $h_j$.  Thus each of the brackets in (\ref{brackM}) uniquely
defines a Poisson bracket on $M_I$ which is rational if written in terms of either
matrix entries of elements of $M_I$ or in terms of the parameters $c_i, d_i$ defined
on $M'_I$ by (\ref{factorI}), (\ref{Cj}). We obtain

\begin{thm}
For any $I$, the restriction of the hierarchy (\ref{lax}) to $M_I \subset \H$
possesses a multi-Hamiltonian structure. Compatible Poisson brackets $\{\ , \ \}^I_k\ 
(k=0,\ldots, n)$ for this structure are obtained as a pull-back of the Poisson brackets
(\ref{brackM}) via the restriction of the map (\ref{weyl}) to $M_I$. The $m$-th flow of the
hierarchy (\ref{lax}) is generated in the Poisson structure $\{\ , \ \}^I_k$ by the
Hamiltonian $H_{m+1-k}(X)$, where
\begin{equation}
H_j =\left \{ \begin{array}{cc} \frac{1}{j} Tr(X^{j}) & j\ne 0\\
\ln\det (X) & j=0\end{array} \right . \ .
\label{hamI}
\end{equation}
\label{thm}
\end{thm}

\medskip

Theorem \ref{thm} provides a uniform way of constructing a multi-Hamiltonian structure
for both standard and relativistic Toda lattices, as well as all lattices $TL_I$.
Furthermore, since the relativistic Toda hierarchy is connected to the Ablowitz-Ladik one
(\cite{al}) via a birational transformation (cf., e.g. \cite{fg}, \cite{kmz}), one can 
use  Theorem \ref{thm} to construct a multi-Hamiltonian structure for the latter hierarchy
too. Note also, that for any $I$, $\{\ , \ \}^I_0$ is the restriction of the 
bracket (\ref{brack1}) to $M_I$

We shall now derive a formula that, for fixed
$\lambda$ and $\mu$, expresses $\{M(\lambda, X), M(\mu, X)\}_k$
in terms of $X$, that agrees with the formula conjectured (and proved for $k=0,1,2$)
in \cite{dam1} for compatible Poisson brackets for the symmetric Toda  lattice.
First, recall the definition of the $R$-matrix associated with the Lax equation
$\ref{lax}$. It is defined by $R(A)= (A)_{\le 0} - (A)_{>0}$ (see, e.g. \cite{rs}).

\begin{prop}
\begin{equation}
\{M(\lambda, X), M(\mu, X)\}_k=
\frac{1}{4} \left\langle X, \left [ R (X^k \nabla_\lambda +  \nabla_\lambda
X^k),\nabla_\mu  \right  ] +   \left [\nabla_\lambda, R (X^k \nabla_\mu +  \nabla_\mu
X^k) \right  ] \right\rangle   \ ,
\label{brackMX}
\end{equation}
where $\nabla_\lambda=\nabla M(\lambda, X)$.
\label{prop3}
\end{prop}

{\bf Proof.} Denote $R_\lambda=(\lambda\one - X)^{-1}$. It follows from (\ref{weyl}) that
$\nabla_\lambda=\nabla M(\lambda, X)=R_\lambda E_{00} R_\lambda$. The following identities
are easily checked:
\begin{equation}
\frac{1}{\lambda -\mu} (R_\lambda - R_\mu) = - R_\lambda R_\mu\ ,  \quad \left [ X,
\nabla_\lambda\right ] = \left [ R_\lambda, E_{00}\right ]\ .
\label{resid}
\end{equation}
Note also that $(\lambda^k M(\lambda))_-=(X^k  R_\lambda e_0,e_0)$. Taking (\ref{resid}) 
into an account, one sees that the second factor in the right-hand side of (\ref{brackM}) is
equal to $(R_\lambda e_0,e_0) (R_\mu e_0,e_0) - (R_\lambda R_\mu e_0,e_0)=
e_0^T R_\lambda (E_{00} -\one) R_\mu e_0$
,while the first factor 
is equal to $((X^k  (R_\lambda  - R_\mu) e_0,e_0) =
e_0^T X^k  (R_\lambda - R_\mu)e_0$. 

Thus, 
\begin{eqnarray}
\nonumber
&\{M(\lambda, X), M(\mu, X)\}_k= Tr\left ( X^k  (R_\lambda  - R_\mu) E_{00}
R_\lambda (E_{00} -\one) R_\mu E_{00}\right ) \\
\nonumber
&= Tr\left ( X^k \nabla_\lambda (E_{00} -\one) R_\mu E_{00} -
E_{00}R_\lambda (E_{00} -\one) \nabla_\mu X^k\right )\\
\nonumber
&= Tr\left ( X^k \nabla_\lambda \left [E_{00}, R_\mu \right ]   E_{00}
-  E_{00} \left [E_{00}, R_\lambda \right ] \nabla_\mu X^k\right )\\
&\stackrel{(\ref{resid})}{=} 
Tr\left ( E_{00} X^k \nabla_\lambda \left [\nabla_\mu , X\right ]   
-  \left [\nabla_\lambda, X \right ] \nabla_\mu X^k E_{00}\right )
\label{inter1}
\end{eqnarray}

Since the Weyl function $M(\lambda, X)$ is invariant under the adjoint action of
a subgroup of $GL(n+1)$ that consists of matrices whose off-diagonal entries in the 
first row and column are zero, one concludes that for any $(n+1)\times(n+1)$ matrix $A$,
that satisfies this property, $Tr\left (\left [\nabla_\lambda, X \right ] A\right )
= \left\langle \left [\nabla_\lambda, X \right ], A\right \rangle = 0$. Then (\ref{inter1})
can be re-written as
\begin{eqnarray}
\nonumber
&\{M(\lambda, X), M(\mu, X)\}_k= \left\langle (X^k \nabla_\lambda)_{>0}, \left [\nabla_\mu ,
X\right ]  \right \rangle  - \left\langle \left [\nabla_\lambda, X \right ] ,(\nabla_\mu
X^k)_{\le 0}\right \rangle \\
&= \left\langle   X,    \left [  (X^k \nabla_\lambda)_{>0}, \nabla_\mu  
\right ] -    \left [ \nabla_\lambda  ,   
(\nabla_\mu X^k)_{\le 0}\right  ]              
\right \rangle \ .
\label{inter2}
\end{eqnarray}

Next, since (\ref{brackM}) defines a skew-symmetric  bracket on $Rat'_{n+1}$ and, hence,
$\{M(\lambda, X), M(\mu, X)\}_k=\frac{1}{2} \left (  \{M(\lambda, X), M(\mu, X)\}_k
- \{M(\mu, X), M(\lambda, X)\}_k \right )$, (\ref{inter2})
implies
\begin{eqnarray}
\nonumber
&\{M(\lambda, X), M(\mu, X)\}_k= \frac{1}{2}
\left\langle   X,    \left [  (X^k \nabla_\lambda)_{>0} - (\nabla_\lambda X^k)_{\le 0} ,
\nabla_\mu  
\right ]  +   \left [ \nabla_\lambda  ,   
(X^k \nabla_\mu)_{>0} - (\nabla_\mu X^k)_{\le 0}\right  ]              
\right \rangle \\
\nonumber
&= \frac{1}{2}
\left\langle   X,    \left [  (X^k \nabla_\lambda + \nabla_\lambda X^k )_{>0} ,
\nabla_\mu  
\right ]  +   \left [ \nabla_\lambda  ,   
(X^k \nabla_\mu + \nabla_\mu X^k)_{>0} \right ] -  \left [  \nabla_\lambda
X^k,\nabla_\mu  \right ] - \left [  \nabla_\lambda ,\nabla_\mu X^k \right ] \right \rangle \\
\nonumber
&= \frac{1}{2}
\left\langle   X,   - \left [  (X^k \nabla_\lambda + \nabla_\lambda X^k )_{\le 0} ,
\nabla_\mu  
\right ]  -  \left [ \nabla_\lambda  ,   
(X^k \nabla_\mu + \nabla_\mu X^k)_{\le} \right ] + \left [ X^k \nabla_\lambda
,\nabla_\mu  \right ] + \left [  \nabla_\lambda , X^k \nabla_\mu  \right ] \right \rangle 
\end{eqnarray}
Taking the average of the last two lines and observing that, due to the Jacobi identity,\newline
$\left\langle   X, \left [ [ X^k,  \nabla_\lambda ]
,\nabla_\mu  \right ]\right \rangle = \left\langle   X, \left [ \nabla_\lambda ,
,[ \nabla_\mu , X^k] \right ]\right \rangle = 0$, one obtains (\ref{brackMX}).
\hfill$\Box$

\medskip
\section{}
In \cite{dam1}, master symmetries were used to establish a multi-Hamiltonian
structure of the symmetric finite non-periodic Toda lattice. (For a definition 
and examples of  master symmetries see, e.g., \cite{master}.)
Namely, a family of vector fields
$Y_m, m\ge 1$  was constructed, that satisfies the properties, that, in the context
of the non-symmetric Toda lattice, can be described as follows.
Let $\L_{Y}$ denote the Lie derivative in the direction of the vector field
$Y$, $\nu_m$ be the Hamiltonian vector field on $Jac$ defined by the right-hand side
of (\ref{lax}) and let functions $H_j$ be defined by (\ref{hamI}). Then  
(i) $[Y_l, Y_m] = (l-m) Y_{l+m}$ ; (ii) $\L_{Y_l} H_j = (l+j) H_{l+j}$;
(iii) $[Y_l, \nu_m] = (m-1) \nu_{m+l}$ and (iv) Lie derivatives $ \L_{Y_l} \{\ ,\ \}_0$
of the Poisson bracket obtained as a restriction of the bracket (\ref{brack1}) to $Jac$
are compatible Poisson brackets on $Jac$. For $l=1,2$, $Y_l$ was defined via a differential
equation of the form
\begin{equation}
\dot X = X^{l+1} + [ X, B_l(X)]\ ,
\label{noniso}
\end{equation}
where an auxiliary matrix $B_l(X)$ was chosen in such a way, that the right-hand side
of (\ref{noniso}) is tridiagonal (such choice is not unique). For $l>2$, vector fields
$Y_l$ were defined recursively as $Y_l = \frac{1}{l-2} [Y_1, Y_{l-1}]$. Various extensions
of the results of \cite{dam1} can be found in \cite{dam2}.

Using results from the previous section, we shall give explicit formulae for master
symmetries $Y_l$. For all $l$, they will be described by nonisospectral equations of
the form (\ref{noniso}).

First, we need to modify the results of \cite{bgs}, \cite{bs} to describe 
nonisospectral flows on $Jac$. Recall that for any $X\in Jac, 
\lambda \in \mathbb C$ there exist uniquely defined vectors 
$P(\lambda)=(p_i(\lambda))_{i=0}^n, 
{\tilde P}(\lambda)=({\tilde p}_i(\lambda))_{i=0}^n$, such that
$p_0(\lambda)={\tilde p}_0(\lambda)=1$ and 
\begin{equation}
X P(\lambda) = \lambda P(\lambda) - p_{n+1}(\lambda) e_n, \quad {\tilde P}(\lambda)^T  X = \lambda {\tilde P}(\lambda)^T - {\tilde p}_{n+1} (\lambda)e_n^T\ .
\label{evp}
\end{equation}
Necessarily, $p_i(\lambda), {\tilde p}_i(\lambda), \ i=1,\ldots, n+1$ are polynomials of degree $i$ (moreover, 
${\tilde p}_i(\lambda)$ are monic). Furthermore, 
$p_{n+1}(\lambda)$ is equal to and ${\tilde p}_{n+1} (\lambda)$ is a scalar
multiple of the characteristic polynomial of $X$. Consequently,
\begin{equation}
\frac{d}{d \lambda}P(\lambda)= \D  P(\lambda), \quad 
\frac{d}{d \lambda} {\tilde P}(\lambda)^T = {\tilde P}(\lambda)^T \tilde \D,
\label{Drl}
\end{equation}
where $\D$ (resp. $\tilde \D$) is a uniquely defined strictly lower (resp. upper)
triangular matrix independent of $\lambda$. Then a differentiation of equalities (\ref{evp}) w.r.t. $\lambda$ leads to the following relations
\begin{equation}
\left [ X, \D \right ] = \one + e_n v^T, \quad 
\left [ X, \tilde \D \right ] = -\one +  \tilde v e_n^T \ ,
\label{string}
\end{equation}
where $v_r, v_l$ are vectors that depend on $X$.

\medskip
\noindent{\bf Remark.} It is worth mentioning that pairs $( X, - \D )$ and
 $X, \tilde \D$ belong to a class  matrix pairs $(X, Z)$ satisfying a
condition rank$( [X, Z] + \one ) =1$. This class plays an important role
in study of classical and quantum solvable models, see, e.g.
\cite{wilson}, where it was studied in connection with the Calogero-Moser
model, and \cite{kasman} where it was used to derive a discrete
time integrable system for the energies certain solvable quantum models.

For any polynomial $Q(\lambda)=\sum_{k=0}^{m}Q_k \lambda^k $, consider 
now a differential equation
\begin{equation}
\dot X = Q(X) + \left [ X, (Q(X) \tilde \D )_{\le 0} - 
(\D Q(X) )_{>0} \right ]\ ,
\label{pureniso}
\end{equation}

\begin{prop}
The vector field defined by (\ref{pureniso}) is tangent to $Jac$.
If $X(t)$ evolves according to (\ref{pureniso}), then the evolution
of functions $\rho_i=\rho_i(X(t)), \lambda_i=\lambda_i(X(t)), \ i=0,\ldots n$ 
and $s_j=s_j(X(t)),\ j=0, \ldots$ defined 
in (\ref{weyl}), (\ref{weylreg}) is given by equations
\begin{equation}
{\dot \rho}_i= 0,\  {\dot \lambda}_i= Q(\lambda_i)\ .
\label{evolrhoniso}
\end{equation}
and 
\begin{equation}
{\dot s}_j = j \sum_{k=0}^{m}Q_k s_{k+j-1}
\label{evolmomniso}
\end{equation}
\label{propniso}
\end{prop}

\noindent{\bf Proof.} To prove the first statement, we have to show
that only diagonal and super-diagonal entries of the right hand side of
(\ref{pureniso}) are nonzero. First note, that 
$ [ X,  (\D Q(X) )_{>0}  ]$ is upper triangular. Therefore, the lower triangular part of of the right hand side of
(\ref{pureniso}) is equal to $ (Q(X) +  [ X, (Q(X) \tilde \D )_{\le 0}  ])_{<0}= (Q(X))_{<0} + ( [ X, Q(X) \tilde \D  ])_{<0}=
(Q(X))_{<0} + ( Q(X) [ X, \tilde \D ])_{<0}=0$ due to the second equality
in (\ref{string}). Similarly, if $A_{>1}$ denotes the part of the matrix
$A$ strictly above the super-diagonal, then  $(Q(X))_{>1} - [ X, (\D Q(X) )_{>0}]_{>1}= (Q(X))_{>1} - ([X, \D] Q(X))_{>1} = 0$ by the first equality
in (\ref{string}). Since $ [ X, (Q(X) \tilde \D )_{\le 0}  ]$ 
is upper Hessenberg, the first statement is proved.

Next, (\ref{pureniso}) implies
\begin{equation}
\dot {X^j} = j X^{j-1} Q(X) + \left [ X^j, (Q(X) \tilde \D )_{\le 0} - 
(\D Q(X) )_{>0} \right ]\ ,
\label{pureniso1}
\end{equation}
Since $\D$ and $\tilde \D$ are, resp., strictly lower and upper triangular, both
$(Q(X) \tilde \D )_{\le 0}$ and $(\D Q(X) )_{>0}$ have zero first row and zero
first column. Thus, it follows from (\ref{pureniso1}) that 
${\dot s}_j = (\dot {X^j} e_0, e_0)= j ( X^{j-1} Q(X)e_0, e_0)$ and (\ref{evolmomniso}) follows.
By (\ref{moment}), $s_j= \sum_{i=0}^n \lambda_i^m  \rho_i(X)$. Then it is easily seen,
that (\ref{evolrhoniso}) is consistent with (\ref{evolmomniso}) and therefore is satisfied
by $\rho_i, \lambda_i$ due to the well-known fact that $\rho_i, \lambda_i$ are determined
uniquely by $s_j (j\ge 0)$.
\hfill$\Box$

Consider now vector fields $\V_l\ (l=1,2,\ldots) $ on $Rat_{n+1}$ defined, in
coordinates $\lambda_i, q_i=q(\lambda_i)$,  by
\begin{equation}
\V_l= \sum_{i=0}^n \lambda^{l+1}_i\frac{\partial}{\partial \lambda_i}\ 
\label{vfrat}
\end{equation}
and let, as before, $H_j= \frac{1}{j}\sum_{l=0}^n \lambda_l^j (j=\pm 1, \pm 2,\ldots )$ and 
$H_0= \sum_{l=0}^n \ln \lambda_l$ and $\{\ ,\ \}_k$ be the Poisson brackets (\ref{brackpq1}), (\ref{brackpq2}).

\begin{prop}
Vector fields $\V_l$ satisfy the following properties:
\begin{itemize}
\item[(i)]  $\L_{\V_l} H_j = (l+j) H_{l+j}$
\item[(ii)] $\L_{\V_l} \{\ ,\ \}_k = (k-l-1) \{\ ,\ \}_{k+l}$
\item[(iii)] $\L_{\V_l} h_j = (j+l-n) h_{j+l} - \sum_{\beta=1}^l H_\beta h_{j+ l- \beta}    $
\end{itemize}
\label{propVl}
\end{prop}

\noindent {\bf Proof.} (i) is obvious. To prove (ii), it suffices to use 
(\ref{bracklambda}), (\ref{bracklambdaq}), 
(\ref{bracklambdaq}) and an identity $(\L_{\V} \{\ ,\ \}) (f, g)=
\L_{\V} \{ f , g \} - \{ \L_{\V}  f , g \} - \{ f , \L_{\V} g \}$.

To prove (iii), recall from (\ref{rat}), (\ref{rqlambda}) that
\begin{equation}
h_j=\sum_{i=0}^n r_i \lambda_i^j=\sum_{i=0}^n \frac{q_i}{p'(\lambda_i)} \lambda_i^j
\label{h_j}
\end{equation}
Since 
\begin{eqnarray}
\nonumber
&\L_{\V_l} \ln p'(\lambda_i)= \sum_{\alpha\ne i}\L_{\V_l} \ln (\lambda_i -\lambda_\alpha)
= \sum_{\alpha\ne i} 
\frac{\lambda^{l+1}_i -\lambda^{l+1}_\alpha}{\lambda_i -\lambda_\alpha}\\
\nonumber
&=\sum_{\alpha\ne i} \sum_{\beta=0}^l \lambda_i^\beta \lambda_\alpha^{l-\beta}
= (n-l) \lambda^{l}_i + \sum_{\beta=1}^l \lambda^{l- \beta}_i H_\beta\ ,
\end{eqnarray}
one obtains from (\ref{vfrat}), (\ref{h_j})
\begin{equation}
\L_{\V_l} h_j = \sum_{i=0}^n r_i ( (j+l-n) \lambda_i^{j+l} - \sum_{\beta=1}^l \lambda^{j+ l- \beta}_i H_\beta) = (j+l-n) h_{j+l} - \sum_{\beta=1}^l H_\beta h_{j+ l- \beta}\ . \hfill \Box
\nonumber 
\end{equation}

We are now ready to prove the following

\begin{thm} Let functions $H_j$ on $Jac$ be defined as in (\ref{hamI}) and, for $l=1, 2, \ldots$, 
let $B_l(X) = (X^{l+1} \tilde \D )_{\le 0} - 
(\D X^{l+1} )_{>0} + \sum_{\beta=1}^l H_\beta (X^{l- \beta})_{\le 0}$.
Then vector fields $Y_l$ on $Jac$ defined by (\ref{noniso}) satisfy
\begin{itemize}
\item[(i)]  $\L_{Y_l} H_j = (l+j) H_{l+j}$
\item[(ii)] $\L_{Y_l} \{\ ,\ \}_k = (k-l-1) \{\ ,\ \}_{k+l}$, where $\{\ ,\ \}_k$ are compatible
Poisson brackets on $Jac$ described by Theorem 3.3.
\end{itemize}
\end{thm}

\noindent{\bf Proof.} All we need to show is that evolution equations induced by vector fields
$Y_l$ on 
the Weyl function $ M(\lambda)$ defined in (\ref{weyl}) coincide 
with evolution equations on $Rat'_{n+1}$ induced by vector fields $\V_l$ on $Rat_{n+1}$ via the
map $m(\lambda) \to M(\lambda)= \frac{m(\lambda)}{h_0}$. To this end, it suffices
to compare equations for Laurent coefficients $s_j$ of $M(\lambda)$. The latter equations
do coincide, which drops out immediately from equations (\ref{evolrho}) and
Propositions \ref{propniso} and \ref{propVl}. \hfill $\Box$

Combined with Proposition \ref{propfg}, Theorem 4.3 allows us  to construct
master symmetries for the Toda flows on $M_I$ for any $I$ and gives
an alternative description of the  multi-Hamiltonian  
structure for integrable lattices $TL_I$.

\end{document}